# Identification of Design Recommendations for Augmented Reality Authors in Corporate Training

Stefan Graser [1], Martin Schrepp [2], Stephan Böhm [1]
[1]Center for Advanced E-Business Studies, RheinMain University of Applied Sciences, Wiesbaden, Germany
[2]SAP SE, Walldorf, Germany
e-mail: {stefan.graser@hs-rm.de, martin.schrepp@sap.com, stephan.boehm@hs-rm.de}

*Abstract*—Innovative technologies, such as Augmented Reality (AR), introduce new interaction paradigms, demanding the identification of software requirements during the software development process. In general, design recommendations are related to this, supporting the design of applications positively and meeting stakeholder needs. However, current research lacks context-specific AR design recommendations. This study addresses this gap by identifying and analyzing practical AR design recommendations relevant to the evaluation phase of the User-Centered Design (UCD) process. We rely on an existing dataset of Mixed Reality (MR) design recommendations. We applied a multi-method approach by (1) extending the dataset with AR-specific recommendations published since 2020, (2) classifying the identified recommendations using a NLP classification approach based on a pre-trained Sentence Transformer model, (3) summarizing the content of all topics, and (4) evaluating their relevance concerning AR in Corporate Training (CT) both based on a qualitative Round Robin approach with five experts. As a result, an updated dataset of 597 practitioner design recommendations, classified into 84 topics, is provided with new insights into their applicability in the context of AR in CT. Based on this, 32 topics with a total of 284 statements were evaluated as relevant for AR in CT. This research directly contributes to the authors' work for extending their AR-specific User Experience (UX) measurement approach, supporting AR authors in targeting the improvement of AR applications for CT scenarios.

## I. INTRODUCTION

Innovative technologies, such as Augmented Reality (AR), create new interaction paradigms. It is essential to identify the application's requirements for developing and designing the respective features. AR authoring refers to the process of creating an AR application through various development steps [1]. In this context, we want to specify the different roles of people in relation to AR authoring, following the differentiation by [1]. In our understanding, AR authors focus on the creation of animations, 3D models, visualizations, and interactive elements (e.g., shadows, textures, color schemes, or sound design) by using authoring tools.

AR authoring can be broadly classified into the interdisciplinary field of software engineering, describing the process of developing software systems [2]. Software requirements elicitation as part of software requirements engineering is the initial step in development, collecting, analyzing, and understanding the relevant requirements and needs of stakeholders [3]–[6].

Previous research analyzed the activities related to the requirements elicitation process [7]–[10] and the effectiveness of requirements elicitation techniques [11]–[14]. Different techniques can be found in the literature. Among the traditional methods, interviews, scenarios, and questionnaires are most commonly applied. For a detailed overview, see [6].

According to [15], requirements are the basis for system design and development. However, not all requirements must be determined for each new application or technology. Using a technology for some time in a certain application domain results in design practices and lessons learned over time, which in turn can be recorded in respective design principles, guidelines, heuristics, or recommendations. These provide an orientation in the form of standards and best practices for system design and development, playing a crucial role in efficiently designing usable interactive technologies in an early stage [16]–[19].

However, applying general recommendations or recommendations from other contexts risks neglecting the new interaction paradigms [18]. Thus, context-specific recommendations are essential for developing and designing new technologies. Current research states a lack of relevant, practical, and applicable design recommendations for AR [1][20], and especially for Corporate Training (CT).

In this article, we aim to identify relevant AR design recommendations for CT, resulting in an updated dataset for AR authors. The study is based on an existing dataset with classified Mixed Reality (MR) design recommendations by [18]. We apply a multi-method approach by enhancing and updating the existing dataset, summarizing the main content of subtopics, and evaluating the relevance of subtopics concerning our research objective. Against this background, we address the following research questions:

- *RQ1:* What practical AR-specific design recommendations have been proposed since 2020?
- *RQ2:* How can the newly identified AR design recommendations be classified?
- *RQ3:* How can the resulting topics be described and communicated?
- *RQ4:* Which topics are relevant for AR authors to improve AR applications in CT?

Based on this, we want to make a further classification of this work, as this has specific relevance for the author's doctoral thesis. Our previous research focused on the User

Experience (UX) evaluation of AR applications in CT. The study results are used to further extend the AR-specific UX measurement approach **UXARcis** [21]. Regarding the User-Centered Design (UCD) process, the research is located in the evaluation phase within the UCD process [22]. Figure 1 shows the UCD process.

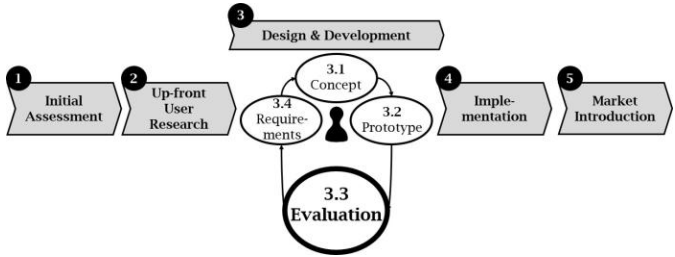

Figure 1. User-Centered Design (UCD) process based on [22]

This means that an AR application no longer has to be developed from scratch, but a functioning AR prototype already exists. In this context, AR authors design the corresponding application features. This focus was considered in the methodological approach of this paper.

The article is structured as follows: Section II introduces the related work to establish a common understanding. Section III explains the multi-method approach of this study, followed by the results in Section IV. A Conclusion is given in Section V, including a Discussion in Section V-A and Limitations in Section V-B. Lastly, our future research is explained in Section VI.

## II. AUGMENTED REALITY DESIGN RECOMMENDATIONS

This Section II introduces the related work for this study. We want to clarify terms and definitions in advance to establish a common understanding. Over the last decades, different terms such as *principle*, *guideline*, or *heuristics* have been established [18][23]. Principles are formulated in general terms [15][24]. Guidelines, on the other hand, are more specific [18]. They can be translated into heuristics, which can be used to evaluate systems [25]. Fu et al. consolidated the following descriptions [23]:

- **Principle**: A fundamental rule or law, derived inductively from extensive experience and/or empirical evidence, which provides design process guidance to increase the chance of reaching a successful solution
- **Guideline**: A context-dependent directive, based on extensive experience and/or empirical evidence, which provides design process direction to increase the chance of reaching a successful solution.
- **Heuristics**: A context-dependent directive, based on intuition, tacit knowledge, or experiential understanding, which provides design process direction to increase the chance of reaching a satisfactory but not necessarily optimal solution.

We will further apply the term *recommendations*, including all three terms, to establish a common understanding for this article.

Context-specific design recommendations are essential as the presentation and interaction of AR differ from other media technologies. Previous research has already worked on this topic, identifying various design recommendations. We mainly refer to the work by [18]. Krauß et al. conducted an extensive literature review, analyzing and summarizing existing approaches, including design recommendations from science and practice up to 2020. In summary, 875 design recommendations for MR applications based on 89 scientific papers and documentation from six industry companies were analyzed. A basic distinction was made between *Scientific Design Recommendations (SDRs)* and *Practitioner Design Recommendations (PDRs)*. This is relevant because, in addition to scientific articles and findings, there are practical recommendations from companies developing the hardware and software concerning AR. The respective statements in both clusters were further analyzed and classified into different main topics, subtopics, and relation to the respective device characteristics (handheld or head-mounted). For details regarding the classification, we refer to [18].

Based on the results, [18] showed that research often adapts traditional (non-spatial) UI principles, such as Nielsen's heuristics [26], without sufficiently addressing MR-specific issues, like ergonomics, spatial interaction, and environment. In contrast, practical recommendations from the industry focus more on MR-specific and practical concerns. Furthermore, SDRs are highly abstract and generic concerning device specification and development, often lacking concrete examples. PDRs are much more detailed, practical, and illustrated with examples to guide application development [18].

To have a positive effect, design recommendations must be seen as essential and valid advice by development teams. But this is often not the case in practice. Design recommendations can be considered as irrelevant and thus be ignored [20] due to the following reasons: Problem of communication, abstraction, research-induced bias [27], ambiguous wording, addressing different target groups with different types of information, and the medium of publication [18].

Based on this, [18] stated relevant implications. Existing design recommendations should be investigated and validated, focusing on the context-specific use of technology. This includes being explicit about the target group, context, and application goals. This should, in turn, be based on transparent, high-quality data and shared practices. Moreover, it is essential to distinguish between research-driven recommendations aimed at enhancing theory (divergent) and practice-driven recommendations supporting system development (convergent) [28]. Existing recommendations should be further structured and classified to facilitate access and better communication channels for practitioners and researchers [18].

Since 2020, little research has been conducted on AR design recommendations. Most papers adopt existing guidelines and apply them to a specific use case. Three articles were found adopting existing recommendations to the area of corporate training (see [29]–[32]).

Agati et al. [29] analyzed existing AR design recommenda-tions for designing AR applications in manual assembly and classified the identified recommendations into the four groups *usability*, *cognitive*, *ergonomics*, and *corporate-related* based on their similarity. Haegle et al. [30] present a set of design recommendations concerning AR assistance in manufacturing. The authors mapped existing recommendations from the literature to the identified challenges in the field of manufacturing machinery.

Only two articles could be identified proposing new guidelines. Chen et al. [32] used public online videos as a basis to identify design patterns, from which they derived new design recommendations for AR-based assembly instructions. Jeffri and Rambli [31] examined the existing types of visual features implemented in AR applications for manual assembly. Based on these, the authors present interface recommendations.

Besides these, no further relevant approaches regarding AR design recommendations were conducted. Our study approach, based on [18], is explained below.

### III. METHODOLOGICAL APPROACH

We applied a multi-method approach containing the four research steps in relation to the research questions:

1) **Identification of new AR design recommendations** based on a Review.
2) **Semantic topic classification of new AR design recommendations** based on an NLP approach.
3) **Content summarization of AR design recommendations topic** based on a qualitative Round Robin approach with AR authors.
4) **Evaluation of relevant topics concerning AR in CT** by AR authors.

The current state of existing design recommendations for MR applications, as presented in [18], provides the basis for this study. We focus on the PDRs, as the SDRs are mostly too generic and, thus, not useful for our research objective [18]. The PDR dataset contains 504 statements, classified into 13 main topics and 84 respective subtopics. Krauß et al. [18] provided their dataset for our research. In the following, we focus on the subtopics as the main topics are too broad. For simplicity, we will only use the term *topic* instead of subtopic in the following. This relates to the subtopics by [18]. An example topic is shown in the Appendix A.

In the first step, we identified new practical AR design recommendations developed since 2020, enhancing the initial dataset of [18] regarding our research objective. The initial dataset by [18] referred to MR and, therefore, also covers other types such as VR based on the reality-virtuality continuum by [33]. In the context of this study, we exclusively focus on AR in relation to the research objective. For this, we followed the approach by [18], analyzing market-leading AR-related companies, including Apple [34], Google [35], Microsoft [36], Magic Leap [37], IBM, and Spark AR [38]. We examined the respective developer documentation containing the design recommendations. However, the developer documentation covering many aspects of development is usually very extensive. The relevant practical design recommendations are typically listed as *best practices*. Hence, we focus our search on the respective statements in relation to *best practices*. Moreover, only design recommendations relevant to AR authors in the respective phase of UCD (see I) were included. Lastly, we also identified practical AR design recommendations from the Nielsen & Norman group during our search [39]. Due to the popularity of the Nielsen heuristics and their inclusion in the SDR by [18], we also considered these. To sum up, we focused on AR-specific statements for best practices within the developer documentation of the market-leading AR-related companies proposed since 2020.

In the second step, we used natural language processing (NLP) to analyze Semantic Textual Similarity (STS) as a common approach for text classification [40], aiming to align the identified statements with existing topics semantically. To achieve this, we applied a pre-trained Sentence Transformer model (SBERT) to analyze STS between each identified statement and the existing topics and their classified statements. This technique enables a fine-grained semantic comparison by transforming textual inputs into dense vector embeddings [41]–[45].

The SBERT is based on the BERT network. BERT is a pre-trained transformer network [46], setting a new benchmark for various NLP tasks [47], indicating the best results for text similarity tasks [48]. The SBERT by [45] enhances the original BERT model using siamese and triplet networks. This enables the application of the SBERT on common STS tasks, such as clustering or text classification [45]. Moreover, previous research showed high potential of applying NLP techniques in UX research activities [49]–[51].

For our analysis, we used the **all-mpnet-base-v2** model demonstrating strong performance across various STS tasks. By computing the cosine similarity between the embeddings of each newly identified statement and those representing the existing topics, we could classify the new data points according to their highest semantic proximity. This vector-based comparison facilitates an accurate assignment of semantically similar content, as discussed in prior work by [45][49]. We used the Python module *sentence transformer library*, including the SBERT for operations [45]. We provided the code as a public repository in git for details, transparency, and comprehensibility (see [52]).

We note that LLMs drive the development and state of the art in NLP tasks. However, small models, such as BERT or SBERT, show sufficient performance for our approach in terms of our dataset and objective [53].

Based on the actualized data set, we applied a Round Robin evaluation approach [54] using five domain experts concerning this topic to summarize and evaluate all 84 topics. Concerning [55], five domain experts are a sufficient number of participants for an evaluation. In particular, three experts work as research associates at a university with more than three years of experience researching, developing, and designing AR applications. The fourth expert has worked as a research assistant at a university, gaining experience with AR authoring and heuristics, and is now a senior consultant in UX design and software requirements engineering. The last author is both a research assistant at the Fraunhofer Institute for Computer Graphics Research IGD and self-employed, researching and working on AR and VR with a practical focus. This specific selection of domain experts ensures that all relevant topics of this research regarding AR authoring, UX design, and software requirements are covered. Moreover, both the research and practice perspective is included.

All topics and their included statements were split equally into five lists, each containing 17 topics and respective statements (one list with 16 topics). The numbers (n - n) within Figure 2 represent the topics within each list. The approach followed a five-round evaluation format. In each round, the experts had the task of analyzing all statements within the 17 topics per list and summarizing the relevant content. The lists with the summaries were then passed on to the next expert, who reviewed and extended the previous expert's input. This iterative process was repeated five times, ensuring that every expert contributed to all 84 topics and all five lists. As a result, each author has completed the required task for all topics. The approach is illustrated in Figure 2:

|          | Expert (1) | Expert (2) | Expert (3) | Expert (4) | Expert (5) |
|----------|------------|------------|------------|------------|------------|
| Round (1)| 1 - 17     | 18 - 34    | 35 - 51    | 52 - 68    | 69 - 84    |
| Round (2)| 18 - 34    | 35 - 51    | 52 - 68    | 69 - 84    | 1 - 17     |
| Round (3)| 35 - 51    | 52 - 68    | 69 - 84    | 1 - 17     | 18 - 34    |
| Round (4)| 52 - 68    | 69 - 84    | 1 - 17     | 18 - 34    | 35 - 51    |
| Round (5)| 69 - 84    | 1 - 17     | 18 - 34    | 35 - 51    | 52 - 68    |

Figure 2. Round Robin Evaluation Approach.

Afterwards, all experts evaluated the relevance of each topic for the use of AR in Corporate Training (CT), specifically in the context of the evaluation phase of the UCD process, where a functional AR prototype already exists. This was done because some topics, e.g., hardware and software compatibility, may be irrelevant, as they pertain more to early development rather than to the refinement of working prototypes.

IV. STUDY RESULTS

Section IV provides the results concerning the four methodological steps. In Section IV-A, we indicate the identified AR design recommendations developed since 2020, followed by their subtopic classification based on the STS in Section IV-B. In Section IV-C, we exemplarily illustrate the subtopic summarization. Lastly, we present the relevant subtopics for AR in CT in Section IV-D. Please note that not all results can be presented in detail due to paper restrictions. We refer to the authors' additional resources for detailed insights (see [52]).

A. Identification AR Design Guidelines

In summary, we elicited 93 new design recommendations specifically referring to AR. In particular, 10 statements were proposed by Nielsen & Norman group [39], whereas Magic Leap provided 83 statements [37]. In particular, Magic Leap proposed five categories regarding best practices for AR. However, three of them do not apply to our research objective because they are too technical or not relevant to the phase of the UCD process. Thus, two categories with 83 statements relevant for AR authors were applied (*Audio Guideline* and *Comfort and Content Placement*) [37].

Google [35], Apple [34], and Microsoft [36] did not publish any further or updated existing guidelines. IBM is not available anymore. This also applies to meta, as meta spark was shut down at the beginning of 2025 [38]. This results in a total of 597 practitioner design recommendations *(as of: April 2025)*.

B. Classification of AR Design Recommendations

The 93 identified statements were classified into 26 topics, grouped under nine main categories. The cosine similarity values range between 0.26 and 0.69. The detailed classification is illustrated in Table I. No fixed cosine similarity threshold was applied, as there is no universally accepted cut-off value that defines semantic similarity in sentence embeddings. Instead, each statement was assigned to the topic with the highest similarity score compared to all other clusters. This ensures that every statement is classified based on its relative semantic proximity, aligning with established practices in clustering and semantic similarity analysis.

Most topics were assigned one, two, or three statements, whereas more statements were assigned to the topics *Audio Feedback* (n = 8), *Audio* (n = 25), and *Content Placement* (n = 9). This is consistent with the description by [37], as the main categories are defined as *Audio Guidelines* and *Comfort and Content Placement*, from which the identified statements are taken.

C. Summarization of Topics Content and Meaning

All topics and respective statements were analyzed, and their content was summarized. In the following, we include the resulting summarization of our exemplary topic (see III). For the comprehensive data, including all descriptions, we refer to our research report [56].

> **Topic**: Consistency
> **Topic summarization**: *This topic is about making your app feel familiar, safe, and easy to use. It includes using standard icons, common interaction patterns, and consistent visuals so users know what to expect. Avoid making people learn new ways to do simple things when familiar ones work just fine.*

TABLE I. STATEMENT CLASSIFICATION BASED ON STS EVALUATION RESULTS.

| Main topic | Topic | Statements | Cosine value |
|---|---|---|---|
| Guidance | Instructions | 1 | 0.41 |
| Input Modalities | Fitt's Law for Touch Interaction | 1 | 0.29 |
| Interactivity | Object Placement | 1 | 0.50 |
| Interactivity | Animations | 2 | 0.38; 0.47 |
| Interactivity | Content SpawnMmechanic | 1 | 0.45 |
| Controls | Control Placement in Screen Space | 1 | 0.35 |
| Multi-User Experience | Shared spaces | 2 | 0.33; 0.38 |
| Design Principles | Customization | 2 | 0.43; 0.47 |
| Design Principles | Law of Practice | 3 | 0.34 - 0.43 |
| Design Principles | Inform about Waiting Time | 1 | 0.48 |
| Design Principles | Information revealing | 1 | 0.43 |
| Design Principles | Accessibility (visuals) | 1 | 0.37 |
| Design Principles | Ergonomics (avoid muscle fatigue) | 2 | 0.42; 0.51 |
| Design Principles | Ergonomics (avoid head & neck fatigue) | 2 | 0.31; 0.33 |
| Technical Requirements | System Architecture | 1 | 0.49 |
| Technical Requirements | Performance | 1 | 0.26 |
| Technical Requirements | Hardware Properties | 1 | 0.30 |
| Feedback | Haptic Feedback (phones) | 1 | 0.52 |
| Feedback | Audio Feedback | 8 | 0.42 - 0.58 |
| Feedback | Feedback | 1 | 0.42 |
| Feedback | Audio | 25 | 0.26 - 0.60 |
| Feedback | Notifications | 3 | 0.40 - 0.46 |
| Spatial Design | FOV | 1 | 0.35 |
| Spatial Design | Content Placement | 9 | 0.28 - 0.46 |
| Spatial Design | Headlocked Content | 3 | 0.40 - 0.50 |
| Spatial Design | Design spaces | 1 | 0.41 |

### D. Relevance of Topics regarding AR in CT

To select the relevant topics, we refer to [57]. The authors described and investigated the determination of the Content Validity Index (CVI) as a representative indicator of quality. They showed that, in a group of five experts, a CVI of at least **0.78** must be achieved to ensure content quality. For calculation, the number of experts who rated it as relevant is divided by the total number of experts. This means that at least four of the five experts (4/5 = 0.8) must classify the respective topic as relevant to reach the threshold [57]. Thus, we excluded all topics rated as relevant by three or fewer experts. This results in **32** topics, with a total of **284** statements. The topics are illustrated in the following. The full list, including the respective statements and summarizations, is provided in [52].

1) Appropriate interplay of virtual content and physical environments
2) Attention directors
3) Instructions
4) Onboarding
5) Hand & finger gestures
6) Textures - Visual Realism and Appearance of Objects
7) Occlusion
8) Image detection
9) Handling Interruptions / Relocalization
10) Surface Detection
11) Affordance
12) Visual cues for object manipulation
13) Object Placement
14) Object Manipulation
15) Encourage to explore
16) Keep the focus on AR experience, but use 2D-UI On-Screen elements when needed
17) Error prevention & recovery
18) Consider and show User's required Effort
19) Law of practice
20) Inform about Waiting Time
21) Text / Font
22) Accessibility (visuals)
23) Ergonomics (avoid muscle fatigue)
24) Ergonomics (avoid head & neck fatigue)
25) Pause / Breaks
26) Performance
27) Audio Feedback
28) Feedback
29) FOV
30) Content Placement
31) Headlocked content
32) Anchored UI

## V. CONCLUSION

This article extends and specifies previous research on MR design recommendations by [18] regarding AR in CT. We applied a multi-method approach to update the dataset with existing design recommendations and further prepare it for our future research regarding AR in CT. In particular, we identified 93 new AR-specific design recommendations, classified into 26 topics, since 2020. We classified them using an NLP classification approach based on a pre-trained Sentence Transformer model, summarized the content of the topics, and evaluated their relevance to AR in CT using a qualitative Round Robin approach with AR authors. As a result, we provide an actualized dataset with AR design recommendations for AR authors relevant to CT. The dataset consists of statements classified into topics, along with a summary of the topic's content and meaning. In the following, we derive implications and limitations.

### A. Discussion and Implications

All research questions could be answered. Based on the results, we derive relevant implications. AR is a rapidly evolving field, driven by continuous advancements in both hardware and software capabilities, frameworks, and interaction paradigms. As AR technologies mature and diversify, so do users' expectations, behaviors, and needs. This dynamic state affects the requirements for designing and developing such applications. Moreover, developers and designers gather valuable experience as more AR applications are implemented across different domains. This results in new insights, refined methods, and design lessons learned from practice over time. This study bridges the five-year time gap, providing an actualized dataset with AR-specific design recommendations proposed since 2020.

Furthermore, the actualized dataset of AR-specific design recommendations provides several practical implications for AR authors regarding applications in CT. First, the summarized topics and their associated statements serve as direct guidance during the design and implementation of AR applications. By integrating these recommendations, AR authors can proactively address known usability and interaction challenges. Second, the topics and respective statements provide a foundation for structured expert evaluations of existing AR applications. Similar to established methods such as heuristic evaluation or cognitive walkthroughs, the recommendations can be applied as a checklist to review functional prototypes or deployed applications [58]. This enables evaluators to systematically assess whether key design principles are met and generate targeted suggestions for improvement. Thus, the dataset not only supports initial design efforts but also fosters continuous quality assurance and iterative enhancement of AR experiences in practice.

*B. Limitations*

While we successfully answered all research questions, some limitations need to be addressed. We want to note that we only focused on the PDRs by the six market-leading AR-related companies, following the approach proposed by [18]. The majority of newly identified statements result from one company [37]. Both the literature on SDR and other practical guidelines, which are certainly helpful, were included. As another aspect, the SBERT is a pre-trained model based on a general training dataset. No ground truth data for training was applied. Moreover, some of the topics are very similar in meaning. Thus, the STS-based classification may be inaccurate.

Moreover, we want to bring up a critical aspect regarding our provided dataset. Both the updated dataset and the final list of design recommendations for AR in CT are complex to use, as they contain a large number of statements. From a practical perspective, it is almost impossible to work with it, either as a checklist or a basis for the review, nor during the design process. To use them effectively in practice, further steps must be taken. For instance, the topics and the respective statements can be assigned to specific system properties. This allows further classification and specification. We want to state that we are aware of this problem. This work provides the basis and preparation for future research to create useful material for AR authors.

## VI. OUTLOOK & FUTURE RESEARCH

Previous research already stated that UX is shaped in the early design phase [59], [60]. However, simply applying design recommendations can lead to problems, as the use of design recommendations depends on the authors' experience [61]. Furthermore, when designing using design guidelines, the subjective opinions of the authors also play a role. They intend to create a certain experience. However, it cannot be ensured that users will actually perceive the intended experience in this way. This is a fundamental discrepancy already described in early UX research [62].

Therefore, without the use of user evaluations, it remains unclear whether the application is perceived as good or bad by the users. It is crucial to identify the application's deficiencies by involving users. However, it remains unclear which weaknesses can be improved by applying which design recommendations. No concrete connection between specific AR design recommendations and the relevant dimensions of user perception exists.

The result provides the basis for extending our previous research. We aim to combine the relevant AR design recommendation topics with the structure of our proposed AR-specific UX measurement approach **UXARcis** [21]. In particular, we want to examine the relationship between specific topics and the UXARcis measurement dimensions by classifying them. As a result, we aim to bridge the gap between empirical UX evaluation methods and their practical applicability for AR in CT, as this remains a major issue in UX research [63].

ACKNOWLEDGMENT

We want to thank Prof. Dr. Veronika Krauß and her co-authors for sharing their data from the previous research.

## Appendix

*Topic*: Consistency
*Statements*:
- *Be reliable. Use global, consistent interactions to make your app both easy to use and safe to explore.*
- *System icons provide users with immediately recognizable visual queues and information. You can use the following icons are available for use in your apps and experiences.*
- *Use consistent, clear, and meaningful symbols.*
- *Certain Control actions must be familiar, intuitive, and adhere to platform conventions*
- *If you use a control ray for selecting things, make sure all your menus work well with the ray. Your user is likely to be confused if they have to switch to another input mechanism such as swiping on the touchpad.*
- *Don't force users to learn a new pattern specific to your app for common interaction when the standard pattern is sufficient*
- *Use familiar UI patterns. Take advantage of your users' knowledge. If there's a standard UX interaction model for a certain action, such as tapping or dragging, use it! You won't have to teach the user a whole new way to perform simple tasks, and you can dive right into the important part of your experience.*
- *Aim for visual consistency. The visuals used for instructions, surface detection, and within the experience itself should share a single consistent look. Aim for visual harmony in all parts of your experience*